\documentclass[a4paper,fleqn]{cas-dc}

\usepackage[numbers]{natbib}
\usepackage{helvet}
\usepackage{xcolor}
\usepackage[normalem]{ulem}

\renewcommand{\baselinestretch}{1} 
\newcommand{\angstrom}{\textup{\AA}}

\newcommand{\tens}[1]{%
  \mathbin{\mathop{\otimes}\limits_{#1}}%
}

\def\tsc#1{\csdef{#1}{\textsc{\lowercase{#1}}\xspace}}
\tsc{WGM}
\tsc{QE}
\tsc{EP}
\tsc{PMS}
\tsc{BEC}
\tsc{DE}

\begin{document}
\let\WriteBookmarks\relax
\def\floatpagepagefraction{1}
\def\textpagefraction{.001}
\shorttitle{\textit{Ab initio} informed electronic stopping models for light ion propagation in metals}
\shortauthors{E. Ponomareva et~al.}

\title [mode = title]{\textit{Ab initio} informed electronic stopping models for light ion propagation in metals}

\author[1]{Evgeniia Ponomareva}
\author[2]{Artur Tamm}
\author[1]{Andrea E. Sand}
\address[1]{Department of Applied Physics, Aalto University, 00076 Espoo, Finland}
\address[2]{Institute of Physics, University of Tartu, 50411 Tartu, Estonia}


\begin{abstract}
Understanding ion-matter interactions at the atomistic level is key to advancing materials for the semiconductor industry, space systems, and nuclear fusion technologies. However, most atomistic frameworks still rely on simplified descriptions of how ions transfer energy to the electronic subsystem, overlooking the sensitivity of this process to the actual ion path. Existing electron-ion interaction models, such as the tensorial unified two-temperature model, were developed to study self-irradiation scenarios, but their suitability for light-ion irradiation remains unexplored. Here, we propose that for light projectiles, stepping back from the tensorial formulation toward a simpler, local model of electronic stopping provides a more efficient and physically transparent trajectory-dependent description. We parameterize and validate both models for hydrogen and helium in tungsten using \textit{ab initio} electronic stopping data and large-scale ion range simulations, benchmarked against existing experimental data. This provides a consistent framework for including nonadiabatic electronic stopping in atomistic simulations of light-ion energy dissipation.

\end{abstract}


\maketitle

\section{Introduction}

Advances in fusion technology hinge on creating materials capable of withstanding the extreme conditions inside nuclear fusion devices \cite{brezinsek2017plasma, pintsuk2019european}. In the harsh reactor environment, plasma-facing components are subjected to intense irradiation by hydrogen isotopes and impurities, leading to sputtering, material erosion and redeposition elsewhere in the reactor \cite{das2019recent}. Gaining a comprehensive understanding of the interactions between light particles and potential reactor materials is crucial for the effective design and operation of fusion systems.

Modern simulation tools play an important role in analyzing and quantifying radiation effects induced by high particle fluxes \cite{trachenko2012modeling, nordlund2018primary, chen2021overview}. The complexity of radiation damage, which spans various scales over space and time, requires balancing computational efficiency with accuracy. At the microscopic level, damage begins with the impact of energetic particles on a material surface, resulting in atomic displacements and near-surface structural modifications. This initial phase of damage sets the stage for subsequent phenomena, which are analyzed using larger-scale models. These comprehensive models rely on parameters derived from primary radiation effects (e.g. defect production rates, sputtering yields, implantation profiles) to simulate long-term microstructural evolution and predict material lifetimes in high-radiation environments \cite{odette2001multiscale, wirth2004multiscale, nordlund2019historical}.

To accurately describe ion transport and near-surface irradiation phenomena, it is crucial to accurately account for how particles transfer energy through their interactions with electrons and atomic nuclei. The retarding force experienced by a particle due to this energy transfer is quantified by stopping power, expressed as an average kinetic energy lost per unit of path length, $S\equiv-dE/dx=S_\text{e} + S_\text{n}$, where $S_\text{e}$ and $S_\text{n}$ are electronic and nuclear components, respectively. At the ion energies typically encountered at plasma-facing surfaces (from hundreds of eV to a few keV), electronic stopping $S_\text{e}$ plays an important role, especially for light projectiles ($Z$ $\leq$ 2). Even at sub-keV energies, the electronic stopping contribution surpasses the nuclear one, which contrasts with the behavior of heavier ions. Accurate quantification of the energy transfer to electrons in this regime is, therefore, required not only for the predictive modeling of fusion-relevant materials but also for the reliable interpretation of low-energy ion-beam experiments \cite{cushman2015low, rubel2023accelerator, wolf2024experimental} used to characterize near-surface composition, erosion, and implantation.

To capture different irradiation-driven atomic-scale processes, molecular dynamics (MD) simulations are often employed. A common approach to incorporate electronic effects into MD involves introducing a velocity-dependent friction force acting on the ions into the equation of motion \cite{finnis1991thermal, nordlund1998defect, gao1998effects}. This approach relies on tabulated friction coefficients, derived from empirical models or \textit{ab initio} calculations, and captures the energy transfer from fast ions to the electronic system. Therefore, it is computationally inexpensive but neglects any feedback from the evolving electronic environment. Over longer timescales, as the lattice and electronic system approach thermal equilibrium, energy exchange between them becomes bidirectional. This regime, governed by electron-phonon coupling, requires to go beyond the Born-Oppenheimer approximation to account for the dynamic interplay between atomic motion and electronic excitations.

To address this, hybrid models such as the two-temperature molecular dynamics (TTMD) framework have been developed, in which ions interact with an electronic heat bath, and the electronic energy evolves according to a coupled heat equation \cite{duffy2006including, zarkadoula2015electronic}. TTMD and its subsequent extensions \cite{zarkadoula2015electronic, zarkadoula2016effects, jarrin2020parametric, wu2021md} have become the standard framework for simulating non-adiabatic energy transfer in radiation damage studies. In this framework, the effective friction coefficient $\beta$ is typically treated as a spatially uniform parameter. However, \textit{ab initio} studies have shown that the rate of electronic energy loss can vary with the local electron density and atomic structure \cite{caro2015adequacy, tamm2016electron, caro2018local, ponomareva2024local}. Therefore, this constant-friction approach neglects the potential environment-dependent effects that can become significant under certain irradiation conditions.

To include environmental effects, the unified TTM model (UTTM) \cite{caro2019role} was developed as the first openly available computational framework capable of incorporating \textit{ab initio}-derived, locally density-dependent electronic stopping functions. The resulting energy dissipation and phonon lifetimes calculated using this method showed a very good agreement with quantum mechanical calculations \cite{tamm2016electron}. However, this increased accuracy comes at the cost of extensive \textit{ab initio} parameterization, typically performed through time-dependent density functional theory (TDDFT) for each element. So far, such parameterizations have primarily been applied to study collision cascades initiated by energetic primary knock-on atoms in nickel (Ni) and Ni-based alloys \cite{tamm2019role, sand2025electronic}, Si \cite{jarrin2021integration}, and GaAs \cite{teunissen2023effect}, significantly affecting defect production and cascade evolution. 

The UTTM framework is formulated for processes occurring in the bulk, ensuring momentum conservation and a physically consistent coupling to lattice vibrations through a tensorial description of friction and stochastic forces. This tensorial formalism implies that the electronic stopping response of each atomic species is interdependent, and parameterizing the model for one element influences the collective description of a multi-component system. For light ions, whose low mass leads to negligible coupling with lattice vibrations, such a detailed description can be unnecessary and inefficient.

In this work, we propose a more direct and computationally efficient description of electronic energy dissipation based on a scalar friction coefficient that depends explicitly on the local electron density, $\beta(\bar{\rho})$. We assess this model for H and He ions in W by combining \textit{ab initio} calculations with the large-scale ion irradiation code MDRANGE, which enables efficient sampling of thousands of ion trajectories and direct comparison with experimental observables. We compare the performance of the proposed $\beta(\bar{\rho})$-model with UTTM, showing that while UTTM was originally developed for uniform lattice systems, it exhibits some limitations for systems with strong mass asymmetry. By analyzing ion range statistics and validating the models through the available experimental measurements, we show that incorporating local electron density dependence is important to accurately describe light-ion dynamics in metallic lattices.

\section{Parameterization of the electron-ion interaction models}

This section guides the reader through the model parameterization process, beginning with the description of real-time TDDFT calculations of electronic energy losses and ending with the translation of density-dependent energy dissipation into a form compatible with a molecular dynamics code. 



\subsection{Electronic energy loss data preparation: \textit{ab initio} simulations}
\label{Sec2.1}

To quantify the energy transfer from a projectile impacting a target electronic system and its correlation with the local electron density, we utilize the open-source real-time TDDFT code Qb@ll \cite{schleife2014quantum, draeger2017massively}. The adiabatic local-density approximation (ALDA) functional is chosen to describe the exchange-correlation effects. A plane-wave cutoff energy of 150 Ry is determined through a convergence test. The time-dependent Kohn-Sham equations are numerically integrated using a fourth-order Runge-Kutta scheme, with a maximum time step set at 0.01 atomic units (equivalent to 0.24 attoseconds). Our simulations utilize a norm-conserving W20 pseudopotential, accounting for 20 electrons explicitly. The simulation setup was thoroughly described in our work \cite{ponomareva2024local}, where we reported on a strong trajectory dependence of light ions electronic stopping in W. 

We consider four different ion trajectories that enable comprehensive sampling of the local electronic environment: $\langle 100 \rangle$ center channeling, $\langle 100 \rangle$ off-center channeling (where the projectile is displaced by $a_0/4$ from the central axis toward the nearest atomic row), $\langle 110 \rangle$ channeling, and a trajectory passing through a vacancy site. The latter represents a direction along which an ion samples the lowest electron density among the studied cases $-$ including this case generally allows to probe the model behavior in non-ideal atomic environments. For this trajectory, a 3 × 3 × 3 cell is used, while for channeling trajectories the simulation box is elongated in the ion propagation direction, comprising 6 × 3 × 3 unit cells of the body-centered cubic (bcc) lattice with a lattice constant $a_0 = 3.16$ $\text{\AA}$. Finally, we note that an additional pseudo-random trajectory (not presented in the Figures), selected from the set generated using the presampling algorithm described in \cite{ponomareva2024local}, was included in the fitting procedure with equal weight as an auxiliary constraint for the high electron density region. We note, however, that the off-center channeling trajectory already samples a sufficiently broad density range, and the influence of the pseudo-random trajectory on the final parameterizations was found to be minor.

For each trajectory, we output the projectile coordinates at each point along its path and the total energy $E_\text{TDDFT}(r)$. The local electron density 
$\rho(r)$ experienced by the projectile is extracted for each trajectory from the ground-state charge density. Additionally, adiabatic Born-Oppenheimer approximation (BOA) calculations of the energy $E_\text{BOA}(r)$ are performed, involving ground state computations at multiple points along the trajectory. The difference between the TDDFT and BOA energies, $E_\text{TDDFT}(r)-E_\text{BOA}(r)$, quantifies the net energy transferred to the electrons by the projectile that can be compared to the output of MD simulations. In all cases, the projectile velocity is fixed at 0.3 atomic units (a.u.), corresponding to projectile energies of approximately 2.25 keV and 8.93 keV for H and He, respectively. The velocity is constrained to remain constant during the simulation \cite{ponomareva2024local}. Assuming the linear $S_e(v)$ dependence, which generally holds for keV and lower energies \cite{bethe1934stopping} for light ions (up to $\sim$30 keV for H and $\sim$200 keV for He), this velocity ensures a reliable linear extrapolation of results within the low-energy range.

\begin{figure*}[t]
	\centering
	\includegraphics[width=1\textwidth]{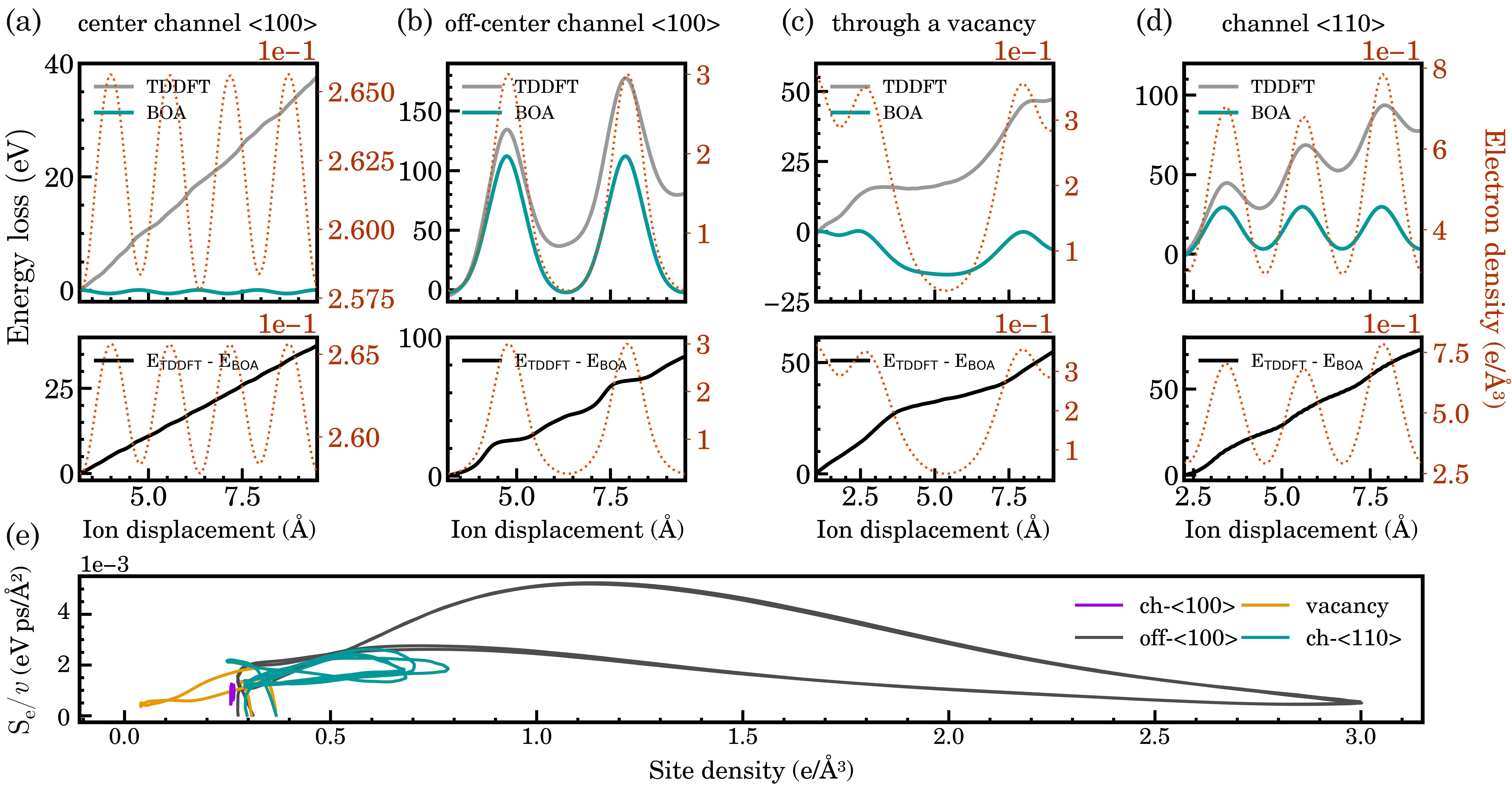}
	\caption{Electronic energy losses of He propagating at $v = 0.3$ a.u. (a) along $\langle 100 \rangle$ center channel, (b) along $\langle 100 \rangle$ off-center channel, (c) through a vacancy, and (d) along $\langle 110 \rangle$ channel in W. In the top panel, the energy profiles obtained from time-dependent (light grey) and ground state calculations (green) are depicted, overlaid with local electron density (orange dotted line with the scale on the right side). The bottom panel shows the difference $E_\text{TDDFT}(r)-E_\text{BOA}(r)$, indicating the net energy transferred to the target electronic system. (e) Electronic energy dissipation strength $S_\text{e}(r)/v$ = $d[E_\text{TDDFT}(r) - E_\text{BOA}(r)]/dr/v$ mapped onto the site electron density for all four trajectories. The multivalued character of the curves reflects the fact that the same local electron density is sampled repeatedly as the ion traverses the periodic crystal environment. This data is used only as a visualization of the raw ab initio density dependence and not directly for parameterization.}
	\label{FIG:2}
\end{figure*}

Figure \ref{FIG:2} shows the results of the real-time TDDFT energy calculations for four selected trajectories of He in W; analogous datasets were obtained for H. The results of the adiabatic calculation (green solid lines) reveal a strong lattice effect, particularly for trajectories that pass through regions with strong electron density variations (Figs. \ref{FIG:2}b–d). Moreover, Fig. \ref{FIG:2}b shows that for close collisions, the peak in the electron density occurs after the energy loss reaches its maximum value. This asymmetry, first identified in \cite{caro2018local}, is attributed to electronic excitation effects. The bottom panel presents the net electronic energy loss obtained by subtracting the adiabatic contribution from the time-dependent results. These difference curves serve as a direct reference for benchmarking MD simulations employing the electron-ion interaction model.

By numerically differentiating these curves with respect to the projectile displacement, we obtain the dissipation strength $d[E_\text{TDDFT}(r)-E_\text{BOA}(r)]/dr$, which can be mapped onto the electron density $\rho(r)$, since both quantities are functions of $r$ \cite{caro2018local}. Figure \ref{FIG:2}e shows the resulting dependence for all four trajectories. For most electron density values, there are at least two corresponding dissipation strength values. This occurs because we mapped the dynamic response of the electronic system onto the static electron density. Although using the ground state density is a significant simplification, it is the only practical approach to construct input functions applicable in atomistic-level simulations. The trajectory dependence observed in Figure \ref{FIG:2}e suggests that the electron-ion interaction should be described by a density-dependent function, rather than by the constant approximation commonly used in atomistic-level simulations. The resulting dependencies can serve as an initial estimate for fitting the density-dependent electronic stopping function, as described in Section 2.3.

\subsection{Integrating electronic effects into atomistic framework}

To translate the \textit{ab initio} data into a function for describing electron-ion interactions in atomistic frameworks, we first need to introduce the corresponding theoretical models. The UTTM model, extensively discussed in \cite{tamm2018langevin, caro2019role}, is implemented as \href{http://github.com/LLNL/USER-EPH}{a fix USER-EPH} in the MD code LAMMPS \cite{thompson2022lammps}. To date, this is the only theoretical framework that accounts for the environment-dependent electronic effects and is parameterized through TDDFT calculations, thereby significantly enhancing the accuracy of atomistic predictions.

The general equation of motion in this model includes the well-known adiabatic term originating from the conservative potential $U_\text{adiab}$, a nonadiabatic drag force, and a stochastic term representing thermal fluctuations:

\begin{equation}
    m_I\frac{\partial\textbf{v}_I}{\partial t} = -\nabla_I U_\text{adiab} - \sum_J  \textsf{B}_{IJ}\textbf{v}_J + \sum_J \textsf{W}_{IJ}\boldsymbol{\xi}_J.
    \label{Eq1}
\end{equation}

\noindent
Unlike the original model proposed by Caro and Victoria in 1989 \cite{caro1989ion} and subsequent expansions of it, Eq. \ref{Eq1} includes nonadiabatic terms expressed as many-body tensors rather than scalars, thereby accounting for spatial correlations between all interacting particles. Tensors $\textsf{B}$ and $\textsf{W}$ are linked through the fluctuation-dissipation theorem, $\textsf{B}_{IJ} = \sum_K  \textsf{W}_{IK}\textsf{W}^T_{JK}$, so that friction and random forces are related to each other. The random vectors $\boldsymbol{\xi}_J$ represent white noise produced by a thermal bath at the local electronic temperature $T_{e}$.

Electronic effects are incorporated into the tensor $\textsf{W}$:

\begin{equation}
  \textsf{W}_{IJ} = \begin{cases}
    - \alpha_J(\bar{\rho}_J)\frac{\rho_I(r_{IJ})}{\bar{\rho}_J} \textbf{e}_{IJ} \tens{} \textbf{e}_{IJ} & (I \neq J), \\
    \alpha_I(\bar{\rho}_I)\sum_{K \neq I}\frac{\rho_K(r_{IK})}{\bar{\rho}_I} \textbf{e}_{IK} \tens{} \textbf{e}_{IK} & (I = J),
  \end{cases}
  \label{Eq2}
\end{equation}
\noindent
where $\textbf{e}_{IJ}$ is the unit vector that joins atoms $I$ and $J$, $\bar{\rho}_I = \sum_{J \neq I} \rho_J(r_{IJ})$ represents the local electron density experienced by atom $I$, while $\rho_I(r_{IJ})$ is the value of the radial electron density function of atom $I$ at distance $r_{IJ}$.

The function $\alpha_I(\bar{\rho}_I)$, referred to as the coupling parameter, bridges the electron-phonon and electronic stopping regimes of ion energy transfer within a material. It must be determined individually for each element in the system through a fitting procedure aimed at minimizing the difference between TDDFT and MD simulation results. Since obtaining the instantaneous electron density from TDDFT is impractical in MD, $\bar{\rho}$ is instead reconstructed from the spherical atomic densities derived from the ground-state DFT calculation. In this work, we calculate these atomic densities for W, H, and He using the pseudopotential generation code Opium \cite{yang2018opium}.

While the full tensorial UTTM model captures spatially correlated dissipation effects, it introduces a level of complexity that may not be essential for light-ion irradiation scenarios, where nonlocal electron-ion correlations are relatively weak. We assume that in such cases, the model can be reduced back to a scalar Langevin form \cite{caro1989ion}:

\begin{equation}
    m_I\frac{\partial\textbf{v}_I}{\partial t} = -\nabla_I U_\text{adiab} - \beta_I(\bar{\rho}_I)\textbf{v}_I + \eta_I, 
    \label{Eq3}
\end{equation}
\noindent
where the friction force $\beta_I(\bar{\rho}_I)\textbf{v}_I$ remains dependent on the local electron density $\bar{\rho}_I$, and $\eta_I$ represents random forces.

In the following section, we attempt to parameterize both models: the full tensorial model, hereafter referred to as UTTM, and its reduced scalar version, denoted as the $\beta(\bar{\rho})$-model. In this context, parameterization means identifying the functional forms of $\alpha(\bar{\rho})$ and $\beta(\bar{\rho})$ that best reproduce the electronic energy losses obtained from TDDFT.

\subsection{Fitting the density-dependent functions}
\label{Sec2.3}

\begin{figure*}[t!]
	\centering
	\includegraphics[width=\textwidth]{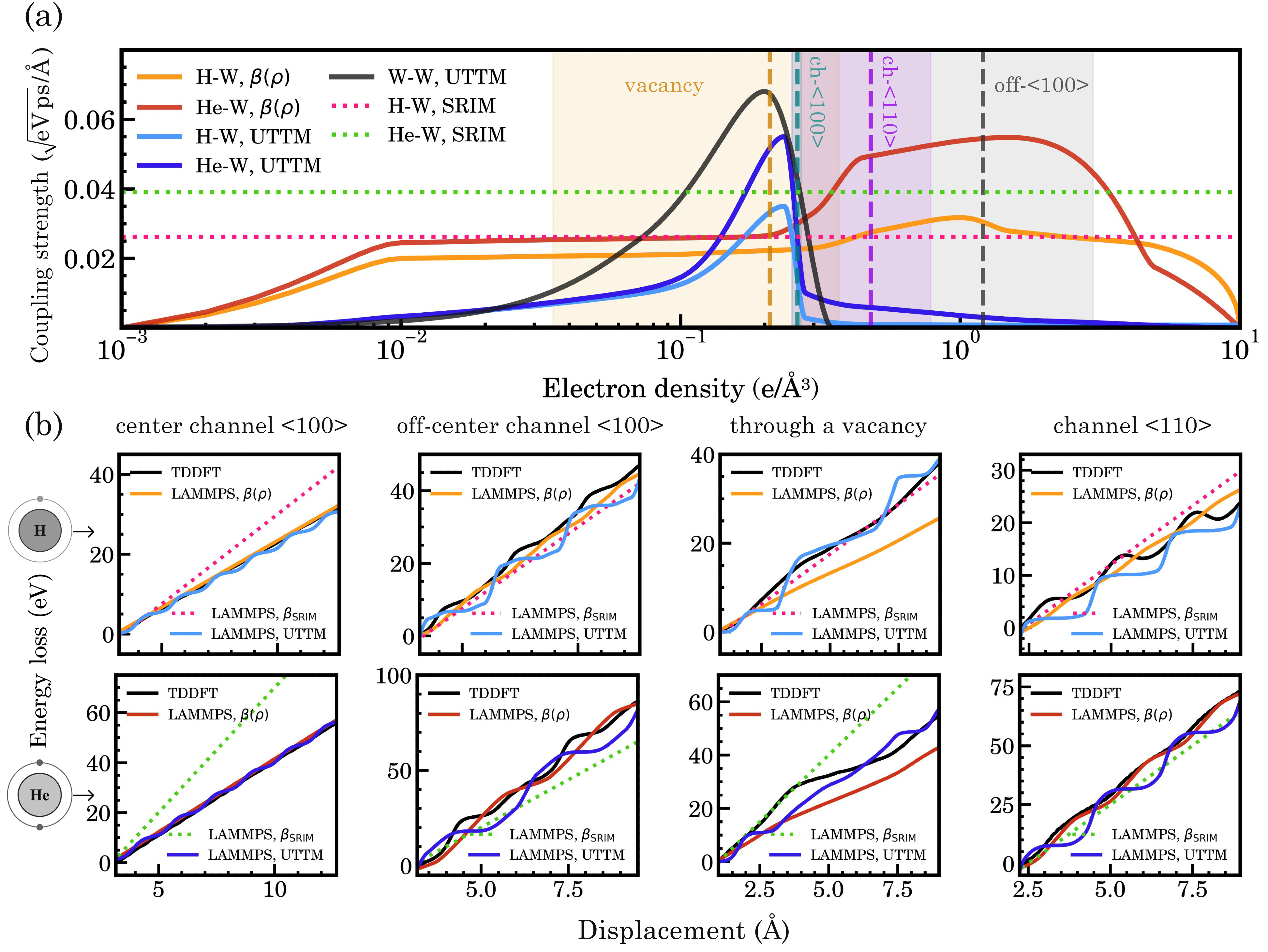}
	\caption{Parameterization of the electron-ion interaction models for H and He propagating in W. (a) Numerically optimized H-W and He-W coupling functions $\alpha(\bar{\rho})$ and $\sqrt{\beta(\bar{\rho})}$ and constant electronic stopping values taken from SRIM for comparison. (b) The dissipated energy, calculated with TDDFT (black), in comparison with those, predicted by MD with UTTM and $\beta(\bar{\rho})$-model (color-coded as the corresponding parameterizations in (a)). Results of MD simulations with constant SRIM-derived values $\beta_{\text{SRIM}}$ are presented for comparison.}
	\label{FIG:3}
\end{figure*}

The fitting procedure aims to determine a density-dependent functional form for the electronic energy loss which, when implemented in MD, reproduces the \textit{ab initio} energy loss data across all selected trajectories. The optimization begins with generating an initial set of test points that serve as a first guess for the density-dependent function. These points are then interpolated to construct a smooth positive function and converted into the format required by the USER-EPH implementation \cite{user_eph}. A friction-only simulation is performed under the same conditions as in the corresponding TDDFT setup, and the resulting energy loss profile is compared to the \textit{ab initio} reference. The objective function, defined as the difference between the TDDFT and MD energy loss curves, $|E_\text{TDDFT}(r) - E_\text{BOA}(r) -E_\text{MD}(r)|$, is minimized iteratively by adjusting the guess points until convergence is achieved. Since the objective function lacks a simple analytical form, we tested several derivative-free optimization algorithms available in the SciPy and NLopt libraries \cite{stevenjohnson}. Among them, the Nelder-Mead \cite{gao2012implementing} and NEWUOA \cite{powell2006newuoa} algorithms demonstrated the most robust convergence and were therefore selected for the fitting process. All functional forms were constrained to remain smooth and positive throughout the sampled density range to preserve physical consistency and transferability. The $\beta(\bar{\rho})$ functions are available in the repository \href{https://version.aalto.fi/gitlab/nume/electron-density-dependent-light-ion-stopping/}{electron-density-dependent-light-ion-stopping}, while $\alpha(\bar{\rho})$ functions can be found in the corresponding UTTM model repository \href{http://github.com/LLNL/USER-EPH}{USER-EPH} \cite{user_eph, light_ions_es}.

Figure \ref{FIG:3}a presents the resulting density-dependent coupling functions together with the constant stopping values obtained from SRIM, which serve here as a reference for the constant-friction approximation. A clear difference between the functional forms obtained for UTTM and for the $\beta(\bar{\rho})$-model can be readily observed. The reduced model only accounts for the energy loss of the projectile and treats the electron-ion coupling as a local process, where the friction term depends only on the projectile velocity and the surrounding target electron density. In contrast, the full tensorial UTTM formulation introduces spatial correlations between all interacting particles, so that the friction acting on one atom depends not only on its own motion but also on the motion of the neighboring atoms through the off-diagonal components of the coupling tensor. This makes the effective interaction between the projectile and the target collective in nature. As a result, UTTM requires defining the coupling function $\alpha(\bar{\rho})$ for every interacting pair in the system, i.e. W-W and ion-W combinations. 

In practice, this formulation makes the parameterization substantially more constrained. In our case, the W-W component must be fitted first with high accuracy, as it determines the collective electronic response of the lattice. Once this is done, however, the flexibility to adjust the ion-target coupling is significantly reduced. This limitation becomes particularly evident for light projectiles, where the target contribution to the overall energy loss can become comparable to the ion's own electronic dissipation. Moreover, since UTTM explicitly accounts for the electronic density of all elements in the system, the effect of the ion electron cloud becomes substantial. For H, convergence could only be achieved by artificially reducing its electronic density by roughly an order of magnitude. Such behavior has not been reported for heavier ions, likely because previous applications of UTTM focused mainly on self-irradiation scenarios or systems with comparable atomic masses \cite{tamm2019role, sand2025electronic, jarrin2021integration, teunissen2023effect}. The present case, therefore, highlights a new regime of mass asymmetry, where the strong imbalance between the light projectile and the heavy target amplifies the nonlocal coupling effects intrinsic to the tensorial formulation.

The dashed vertical lines in Fig. \ref{FIG:3}a mark the average electron density sampled along each trajectory, while the shaded regions indicate the full span of densities encountered by the projectile during its propagation. The widths of these spans differ significantly between trajectories: the $\langle100\rangle$ center channeling path samples a very narrow interval of low density, whereas the off-center channeling ion in the same direction experiences a much broader portion of the electronic environment. It is important to emphasize that these results were obtained within the straight-line trajectory approximation \cite{ponomareva2024local}.

Figure \ref{FIG:3}b shows that the constant friction parameters obtained from SRIM systematically overestimate the energy loss for the $\langle100\rangle$ center channeling trajectory, as SRIM provides an orientation-averaged stopping value and therefore cannot capture the reduced electronic dissipation characteristic of open crystallographic channels. In contrast, the density-dependent parameterizations obtained in this work reproduce the slope of the TDDFT energy loss curves across all periodic directions with good accuracy. The UTTM parameterizations perform consistently well across all four trajectories for both ions, while the $\beta(\bar{\rho})$-model parameterizations show slightly larger deviations for the trajectory passing through a vacancy. However, if we focus only on the actual vacancy region, where the increase is less steep, the local energy loss in the low-density region is well described by $\beta(\bar{\rho})$. Additionally, all fitted functions are constrained to reproduce the slope of the pseudo-random trajectory, mentioned in Sec. 2.1. The coverage of the local environments represented by the trajectories used for the parameterizations is assessed in Appendix B.

\section{Validation of density-dependent functions via statistical ion range simulations}

\subsection{MDRANGE for simulating ion depth profiles}

In this section, we evaluate the adequacy of the obtained parameterizations using statistical ion range simulations. These were performed with the MD-based code MDRANGE \cite{mdrange}, which enables efficient calculation of ion trajectories by employing the recoil interaction approximation (RIA) \cite{nordlund1995molecular}. Using RIA, only the interactions between the projectile and the nearby target atoms are explicitly calculated. The surrounding lattice atoms remain fixed until they come within the cutoff range of the moving ion, at which point they interact via a two-body repulsive potential. This approach omits explicit lattice-lattice dynamics, enabling high statistical accuracy at a greatly reduced computational cost.

The latest version of MDRANGE supports multiple electronic stopping models, including the constant-friction model, UTTM \cite{kiely2026trajectory}, and the $\beta(\bar{\rho})$ model described here, which makes it a suitable platform for testing the developed parameterizations on a statistically representative ensemble of trajectories. For each case, 10,000 ions were initialized at a distance of $z$ = $-$3 Å from the target surface, with $x$ and $y$ positions randomized within one unit cell of the $\langle100\rangle$ surface. The Ziegler-Biersack-Littmark (ZBL) universal repulsive potential \cite{ziegler1985stopping} was used to describe the ion-atom interaction. Thermal displacements corresponding to 300 K were assigned according to the Debye model, using a Debye temperature of 310 K for W \cite{ashcroft1976solid}.

In general, all simulations followed a similar workflow, with variations only in the electronic stopping model or in the geometric setup corresponding to different crystal directions. While setting up channeling directions is straightforward in atomistic simulations, achieving a truly random direction requires suppressing long-range channeling effects \cite{nordlund2016large}. For this purpose, a polycrystalline target was generated using the built-in functionality of MDRANGE. The grains were randomly oriented with an average size of 20 ± 5 Å, effectively minimizing the probability of ion channeling.

The purpose of these simulations is to evaluate the performance of different electronic stopping models in predicting ion range profiles. In the absence of experimental data, we have to select appropriate references. For random directions, SRIM-based electronic stopping provides a suitable benchmark, as it effectively represents the statistical average of multiple ion trajectories. In contrast, for channeling directions, an accurate parameterization is shown to yield a maximum ion range $R_\text{max}$ comparable to that predicted by a constant-friction model using the TDDFT-derived electronic stopping for a given channel \cite{sand2019heavy}. This, however, relies on the assumption that the most deeply penetrating ions remain close to the geometrical center of the channel, where deviations from straight trajectories and local density fluctuations are negligible. The validity of this assumption is examined in Appendix A by analyzing realistic ion trajectories in the $\langle100\rangle$ channel.

\subsection{Ion range calculations for H and He in W}

\begin{figure}[b!]
	\centering
	\includegraphics[width=\columnwidth]{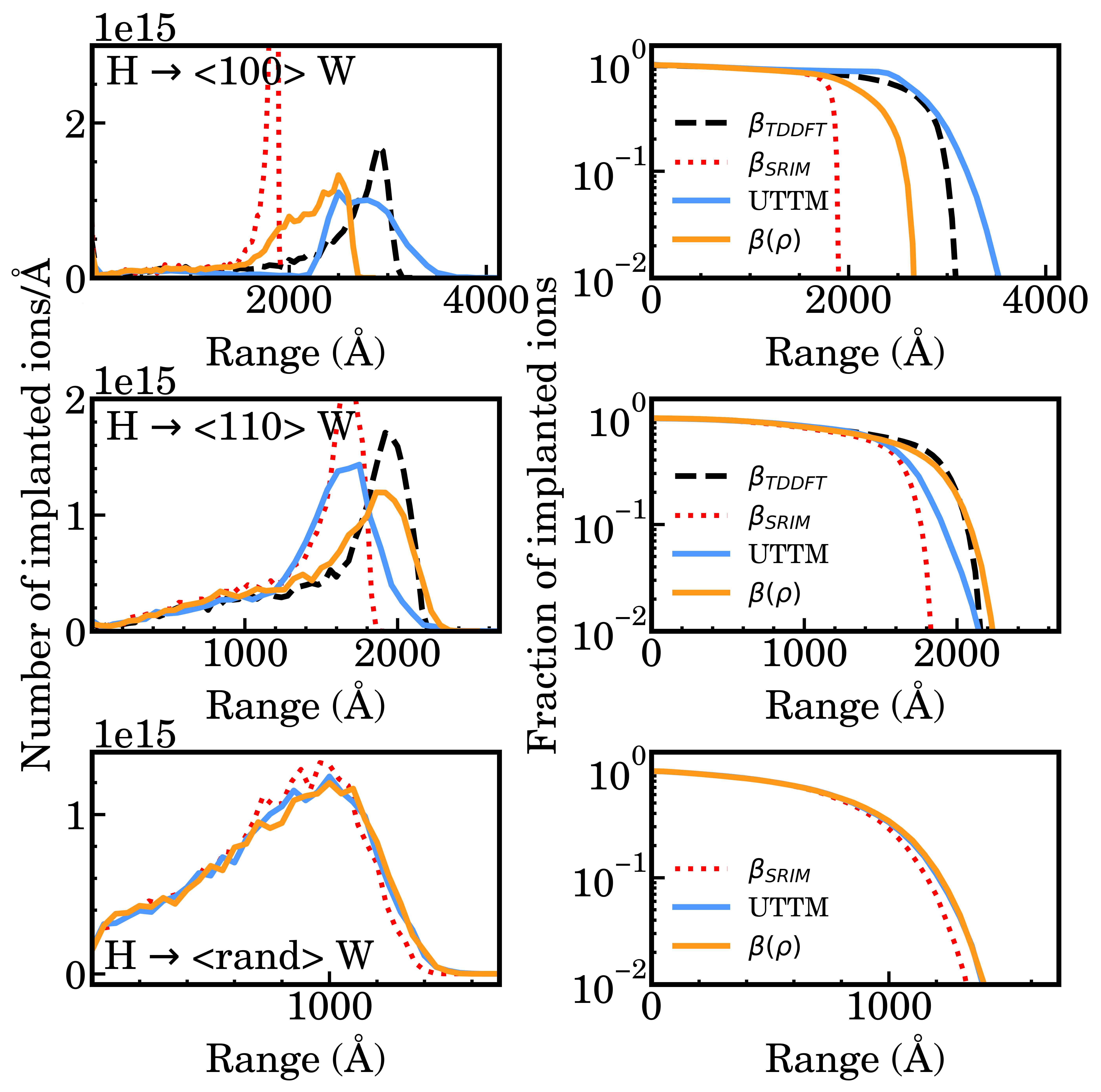}
	\caption{Calculated differential (left) and integral (right) projected range profiles for 10 keV H implanted in different directions in W. Color coding corresponds to the functional forms shown in Fig. \ref{FIG:3}a that served as electronic stopping input for these calculations.}
	\label{FIG:5}
\end{figure}

\begin{figure}[b!]
	\centering
	\includegraphics[width=\columnwidth]{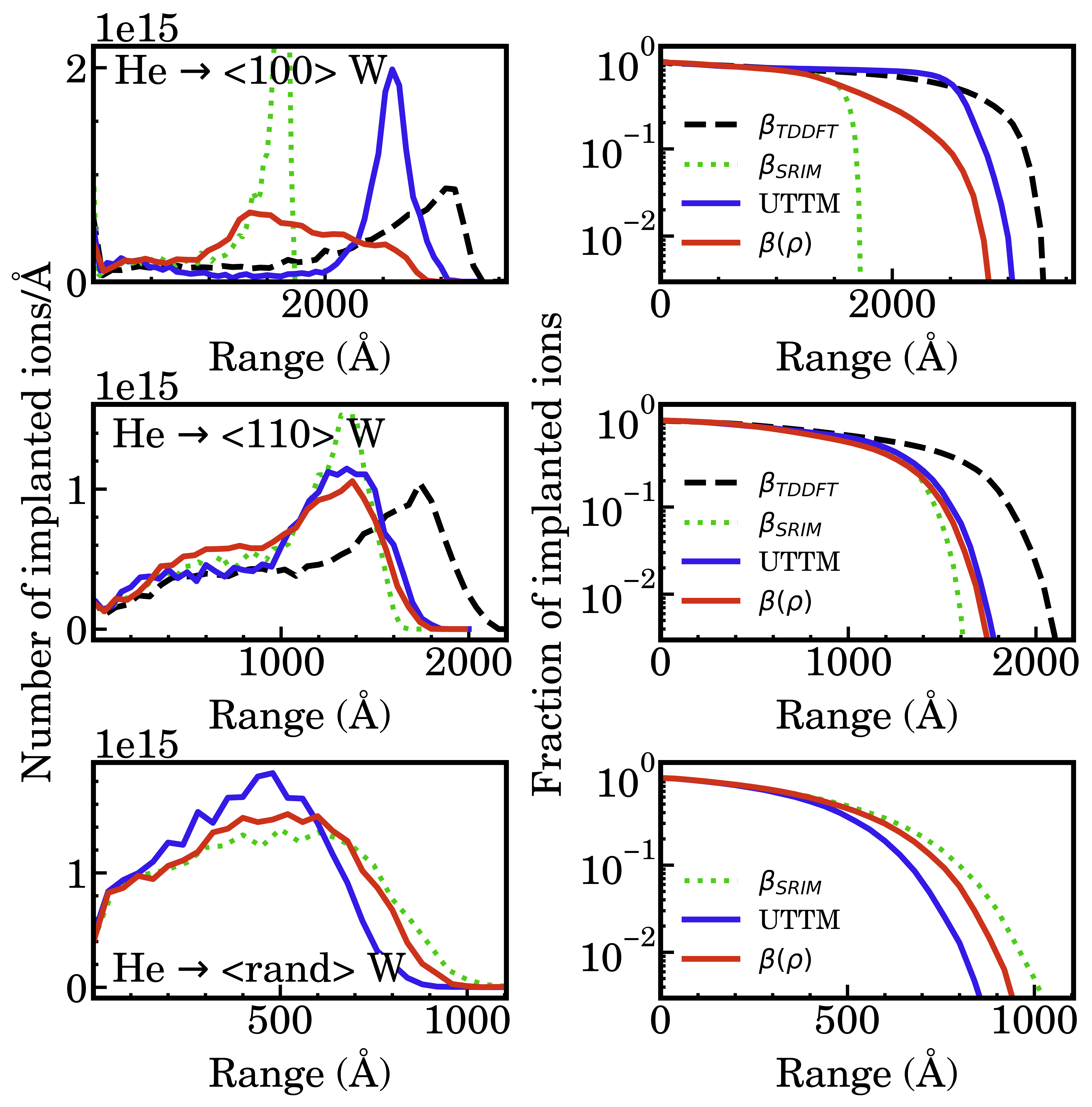}
	\caption{Calculated differential (left) and integral (right) projected range profiles for 10 keV He implanted in different directions in W. Color coding corresponds to the functional forms shown in Fig. \ref{FIG:3}a that served as electronic stopping input for these calculations.}
	\label{FIG:6}
\end{figure}

This section presents the results of projected ion range profile calculations for H (Fig. \ref{FIG:5}) and He (Fig. \ref{FIG:6}) projectiles, obtained using the different electronic stopping models and parameterizations described in Section 2.3. In each figure, the left column shows the differential range distribution, representing the probability density of ions stopping at a given depth, while the right column presents the integral range distribution, indicating the fraction of implanted ions within a certain penetration depth. In the integral range plots, the maximum penetration depth $R_\text{max}$ is identified by the abrupt tail of the distribution. All simulations were conducted with an initial ion energy of 10 keV. 

The comparisons highlight how variations in the friction coefficients and their dependence on local electron density influence both the $R_\text{max}$ and the overall shape of the ion range distributions. For both H and He, all parameterizations reproduce range profiles for a random trajectory consistent with those obtained by using SRIM-based stopping, indicating that the direction-averaged electronic stopping is well captured. However, when ions propagate along low-index crystallographic directions, differences between the models become pronounced. Despite fitting the $\langle100\rangle$ channel trajectory closely for both the UTTM and simplified $\beta(\bar{\rho})$ models (as shown in Fig. \ref{FIG:3}b), the resulting ion ranges diverge significantly.

 In particular for H, the UTTM model predicts an $R_\text{max}$ exceeding that obtained with a constant-friction model with the TDDFT-derived electronic stopping coefficient for this channel. This outcome is surprising, as the constant-friction case represents the idealized center-channeling trajectory, which is expected to yield the longest possible maximum range for this direction due to its low electron-density path. At the same time, the range reduction predicted by the $\beta(\bar{\rho})$ model, relative to the constant-friction prediction, agrees well with the difference of approximately 11\%, estimated in Appendix A. For the $\langle110\rangle$ channel, the peak of the range distribution obtained with the $\beta(\bar{\rho})$ model coincides with that of the TDDFT-based constant-stopping approach, while the UTTM peak aligns with the constant-stopping model using SRIM data. Overall, the peak positions predicted by all models remain close to one another, indicating comparable mean penetration depths within this channel.

For He projectiles, likely due to the smaller ion-target mass asymmetry, all models predict maximum penetration depths that remain within the reference limit set by the constant-friction case for the $\langle100\rangle$ channel. A distinct feature, however, emerges in the UTTM results for both ions: the differential range distributions exhibit a decrease preceding the main peak (around 2000 $\text{\AA}$). In contrast, the simplified $\beta(\bar{\rho})$-model yields a broader distribution with its peak shifted toward shorter penetration depths. This behavior reflects enhanced dechanneling, likely caused by stronger nuclear scattering of He, which increases the model sensitivity to local variations in the friction coefficient. As shown in Fig. \ref{FIG:A1} in Appendix A, the dissipation varies more sharply for He than for H, consistent with the broader spread observed in the corresponding range distributions.

We note that for randomly oriented trajectories, all models yield comparable results, and the averaging over directions effectively suppresses the impact of model variations. Nevertheless, using SRIM-derived friction values for channeling trajectories systematically underestimates the penetration depth, emphasizing that empirical stopping powers fail to capture the anisotropic and density-dependent character of ion motion in crystalline targets.

\section{Comparison with existing experimental data}

As a first validation case, we compare the performance of different electronic stopping models against experimental measurements of the He mean ion ranges in W \cite{kornelsen1980enhanced, wagner1979range}. To account for non-ideal beam conditions and experimental uncertainties in ion incidence, the simulated projectiles were assigned an angular spread of 5$^\circ$ in both polar ($\phi$) and azimuthal ($\theta$) directions. For simulations corresponding to random incidence, the ion azimuthal angle was fixed at 19$^\circ$ with respect to the (100) plane, following the experimental geometry described in Ref.~\cite{kornelsen1980enhanced}.

\begin{figure}[t!]
	\centering
	\includegraphics[width=\columnwidth]{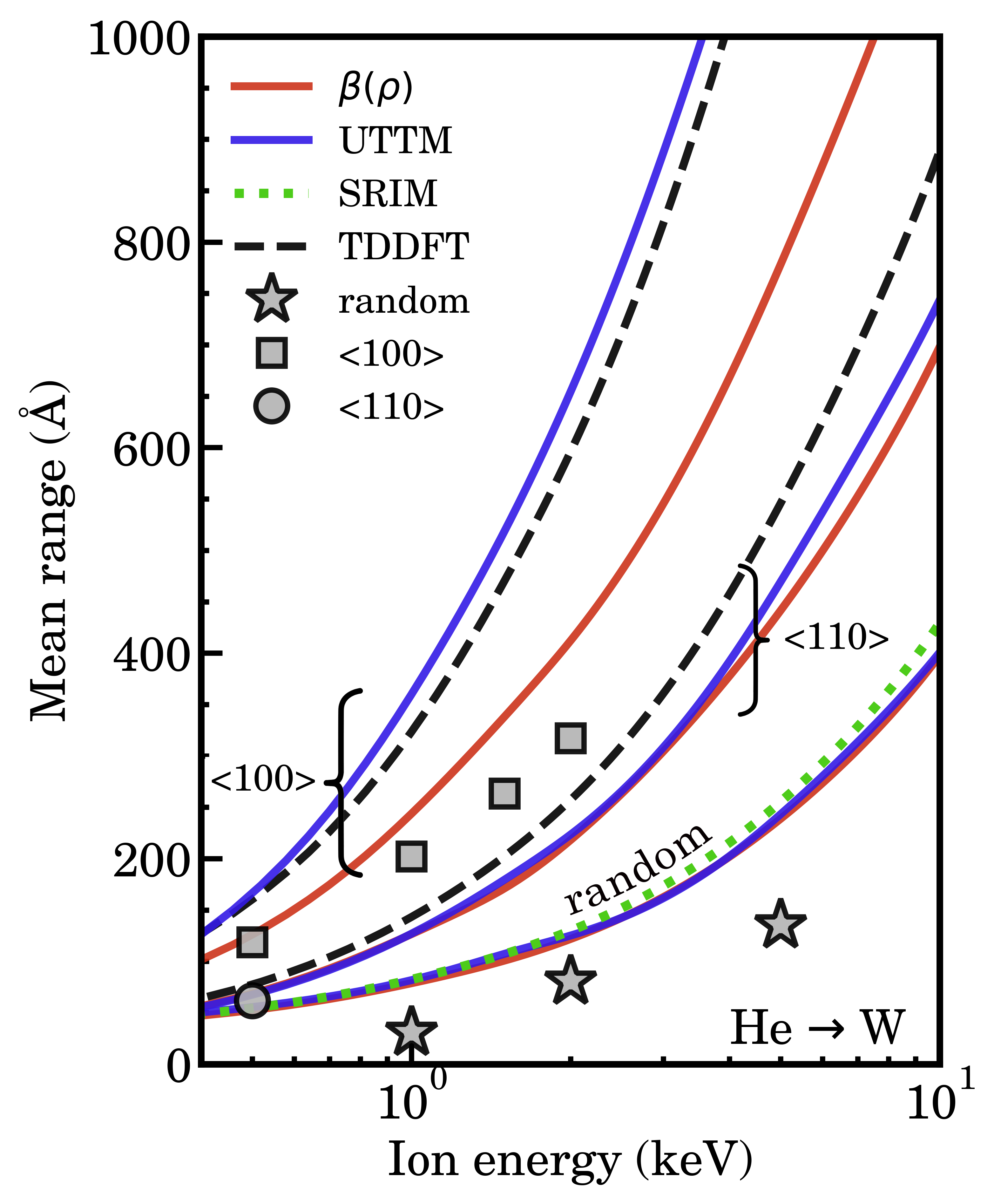}
	\caption{Calculated mean projected range values for 0.5-10 keV He launched in different directions in W. Experimental data (gray symbols) is taken from \cite{kornelsen1980enhanced, wagner1979range}. Color coding corresponds to the functional forms shown in Fig. \ref{FIG:3}a that served as electronic stopping input for these calculations.}
	\label{FIG:7}
\end{figure}

Figure \ref{FIG:7} presents the calculated mean projected ranges for He in W for the $\langle100\rangle$ and $\langle110\rangle$ channeling directions, as well as for random orientations, alongside the corresponding experimental data. The calculations were performed for initial ion energies of 0.5, 1, 1.5, 2, 5, and 10 keV, with the smooth curve in the figure obtained by numerical interpolation of these data points. For both random and $\langle110\rangle$ directions, the trajectory-dependent models produce comparable results that show moderate agreement with the available experimental data. In contrast, the constant-stopping model tends to overestimate the mean penetration depths for channeling directions. For the $\langle100\rangle$ channel, larger discrepancies are observed: although none of the models reproduce all experimental data points, the $\beta(\bar{\rho})$ model consistently predicts lower mean ranges that lie closer to the measured values. Interestingly, the shift of the range profile peak predicted by the $\beta(\bar{\rho})$ model for the $\langle100\rangle$ direction (Fig. \ref{FIG:6}) corresponds to a mean range that aligns more closely with the experimental data, indicating a systematic trend in the model behavior.

To extend the validation of the parameterized electronic stopping models, backscattering energy spectra of deuterium (D) ions in W were simulated and compared with recently obtained experimental data from \cite{shams2024interactions}. Although the parameterizations investigated here were derived for hydrogen, low-energy experimental range or backscattering data for H in W is not available in the literature. We therefore use the D measurements as a proxy, assuming negligible isotope effects. This assumption is supported by recent experimental measurements that show no significant difference between H and D electronic stopping cross sections in W within the experimental uncertainties \cite{shams2023experimental}. The MDRANGE simulations were set to replicate the experimental conditions, including an incident ion energy of 2.66 keV, an exit angle of 51$^\circ$ with an angular spread of 2$^\circ$ \cite{bruckner2020influence}, and a W foil thickness of 98 $\text{\AA}$. To approximate the polycrystalline experimental conditions, we use an amorphous W cell, generated by disordering the crystalline lattice in LAMMPS, representing randomly oriented grains. All spectra were obtained by averaging over a large number of independent ion trajectories (approximately 200,000 impacts per spectrum) to ensure good statistical convergence.

\begin{figure}[t!]
	\centering
	\includegraphics[width=\columnwidth]{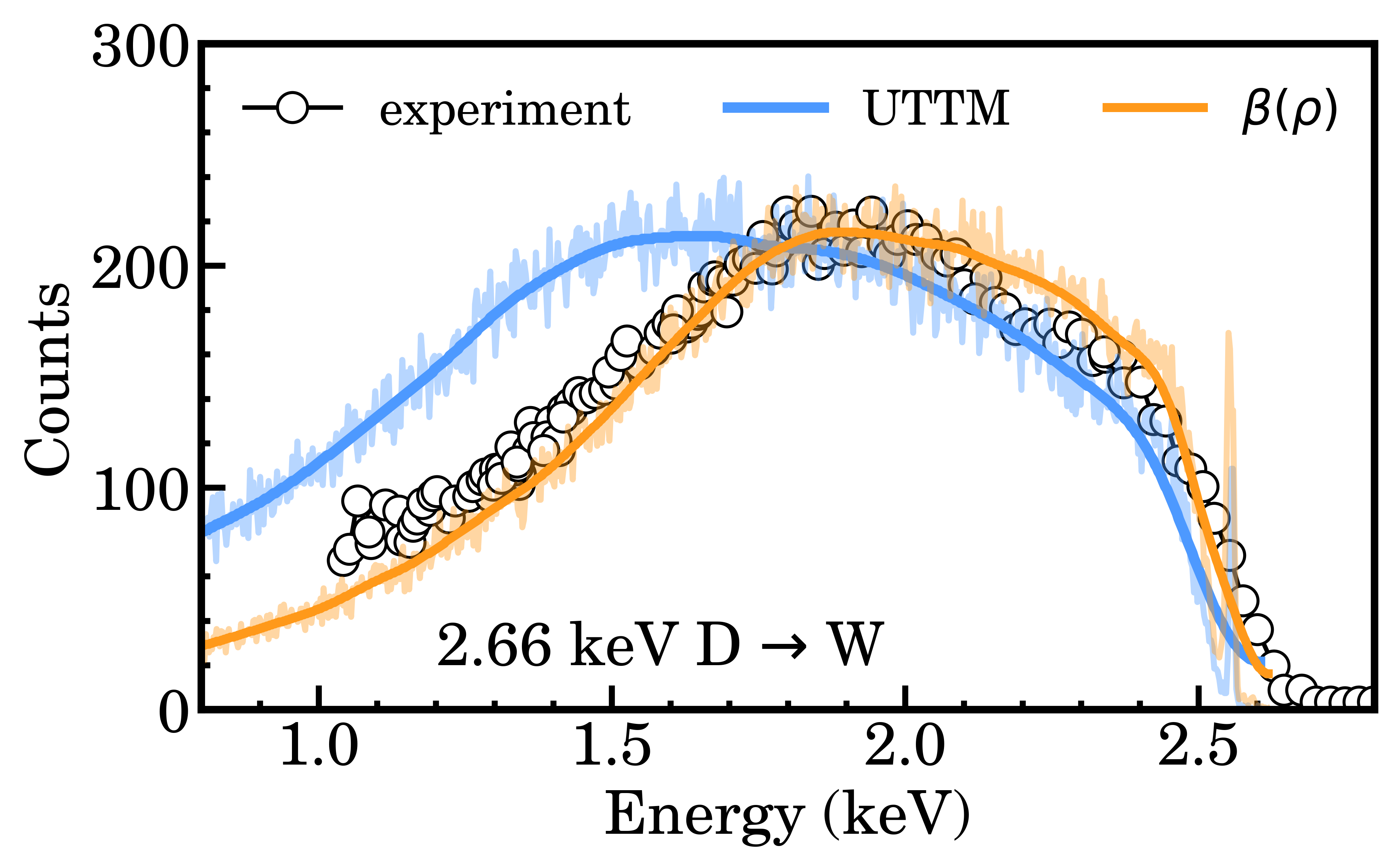}
	\caption{Experimental and calculated spectra of 2.66 keV D ions backscattered from 98 $\text{\AA}$-thick W foil. Experimental data is adapted from \cite{shams2024interactions}. The solid curves represent the backscattering spectra after applying a one-dimensional Gaussian smoothing filter to suppress statistical fluctuations. }
	\label{FIG:8}
\end{figure}

Figure \ref{FIG:8} presents the comparison between the calculated and experimental backscattering spectra. The $\beta(\bar{\rho})$-model reproduces the experimental spectrum shape and the peak position with a high accuracy. In contrast, the spectrum obtained using the UTTM model exhibits a shift of the peak toward lower energies and a slightly broader distribution. In general, this indicates that the ions experience a wider range of energy losses along their trajectories and seem to lose more energy in close approaches. These results are consistent with the trends observed in the analysis of ion ranges and support the conclusion that the reduced $\beta(\bar{\rho})$-model provides a more simple, computationally efficient and physically consistent description of electronic stopping for light ions.

\section{Conclusions and outlook}

In conclusion, this study provides the first systematic evaluation of trajectory-dependent electronic stopping models for light ions in molecular dynamics simulations. Using larger-scale ion range simulations, we demonstrated that fitting electronic energy dissipation to straight-line trajectories alone does not guarantee predictive accuracy in molecular dynamics, where ions explore a broad spectrum of local electron densities. A physically reliable model must therefore respond dynamically to these density variations along the projectile path, capturing the effects of channeling and dechanneling.

The unified two-temperature model (UTTM), while effective for self-irradiation scenarios, was found to overestimate penetration depths in wide channels and incapable of fully capturing ion backscattering properties. To tackle the limitations that seem to arise from the tensorial nature of the model, we employed a simpler local friction $\beta(\bar{\rho})$-model that preserves trajectory dependence while improving stability. By linking the friction force solely to the instantaneous local electron density sampled by the ion, this reduced model reproduces both the magnitude and spatial dependence of electronic dissipation, yielding good agreement with existing experimental mean ranges and backscattering spectra.

As a result, this work establishes a physically grounded and computationally practical approach for modeling electron-ion interactions in atomistic simulations.

\section*{CRediT authorship contribution statement}

\textbf{Evgeniia Ponomareva}: Writing – review \& editing, Writing – original draft, Visualization, Validation, Software, Methodology, Investigation, Formal analysis, Data curation, Conceptualization. \textbf{Artur Tamm}: Writing – review \& editing, Methodology, Supervision. \textbf{Andrea E. Sand}: Writing – review \& editing, Supervision, Project Administration, Conceptualization, Funding Acquisition.

\section*{Declaration of competing interest}

The authors declare that they have no known competing financial interests or personal relationships that could have appeared to influence the work reported in this paper.

\section*{Acknowledgements}

The authors acknowledge the computational resources provided by the Aalto Science-IT project and CSC – IT Center for Science, Finland. This work has been partly supported by the EUROfusion Consortium, funded by the European Union via the Euratom Research and Training Programme (Grant Agreement No. 101052200 — EUROfusion). Views and opinions expressed are however those of the author(s) only and do not necessarily reflect those of the European Union or the European Commission. Neither the European Union nor the European Commission can be held responsible for them. 

The authors also would like to thank Alfredo A. Correa and Glen P. Kiely for valuable discussions.

\section*{Appendix A. Assessing the straight-line trajectory approximation. }

\renewcommand{\thefigure}{A.\arabic{figure}}
\setcounter{figure}{0}

\begin{figure}[t!]
	\includegraphics[width=0.85\columnwidth]{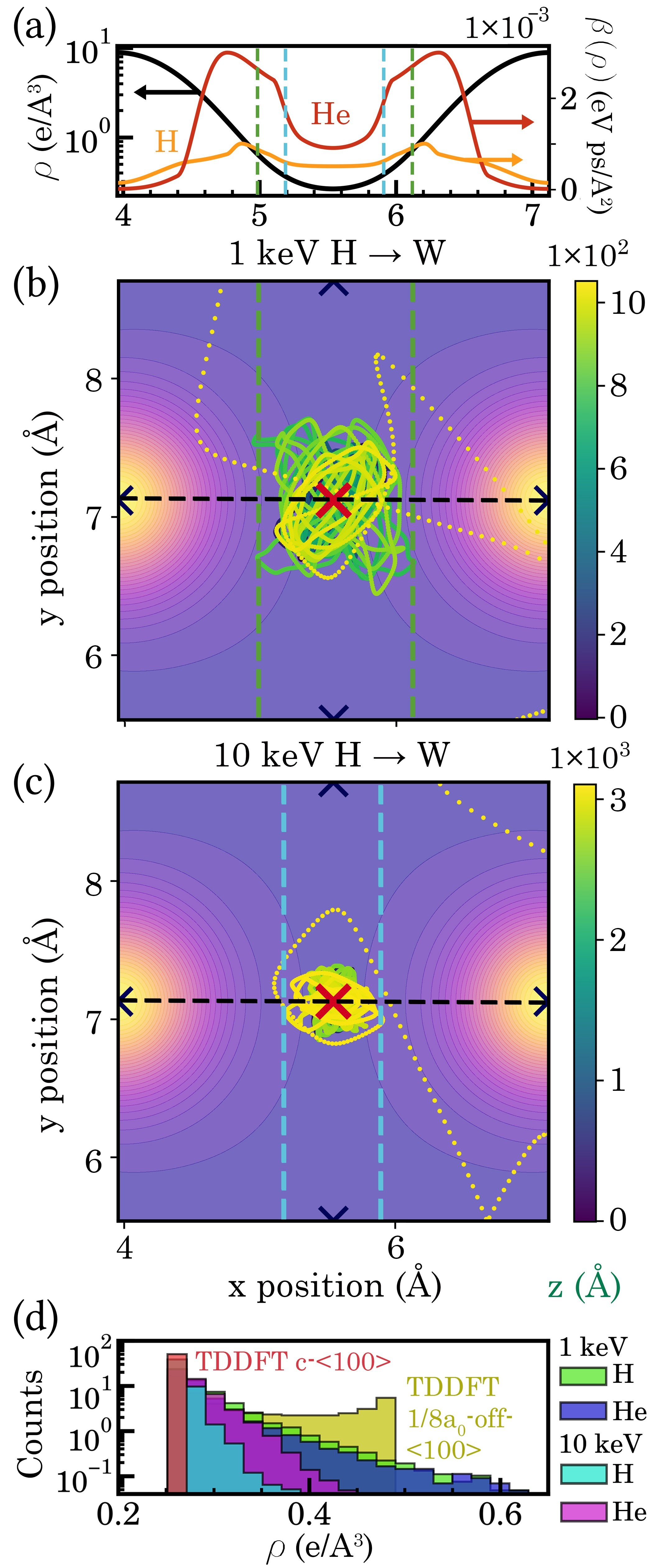}
	\caption{MDRANGE-based analysis of H projectile moving along the $\langle100\rangle$ channel in W. (a) Local electron density and corresponding density-dependent electronic energy dissipation for H and He along the perpendicular line cut indicated by the dashed line in the cross-sectional plots below. (b,c) Projected $x$-$y$ cross-sections of the (b) 1 keV and (c) 10 keV H trajectories. Cyan and green vertical dashed lines indicate the span of transverse ion oscillations around the center of the channel. The color scale on the right indicates the ion $z$ coordinate. (d) Distribution of local electron densities sampled by the most strongly channeled H trajectories, compared with TDDFT reference values obtained for channeling trajectories within the straight-line approximation.}
	\label{FIG:A1}
\end{figure}

Figure \ref{FIG:A1} examines the validity of the straight-line channeling approximation by analyzing realistic propagation of H ions along the $\langle100\rangle$ channel. For computational efficiency, a constant TDDFT-derived stopping value corresponding to this channel was used. Figure \ref{FIG:A1}b and Figure \ref{FIG:A1}c compare the $x$-$y$ cross-sections of deeply penetrating ions ($R > 0.95R_\text{max}$) with initial energies of 1 keV and 10 keV. In agreement with classical channeling theory \cite{lindhard1965influence,bergstrom1968critical}, lower-energy ions exhibit larger transverse oscillations within the channel. Consequently, even strongly channeled ions sample a range of local electron densities rather than remaining at the channel center.

Figure \ref{FIG:A1}a connects these density variations to the density-dependent friction function for both H and He ions. Mapping the oscillation amplitudes onto the fitted $\beta(\bar{\rho})$ curves shows that the effective electronic stopping differs from the ideal center-channeling value. For example, in the case of 10 keV H, the variation in $\beta(\bar{\rho})$ over the oscillation amplitude corresponds to a change of approximately $10^{-4}$~eV ps/\AA$^{2}$. Approximating the transverse motion as sinusoidal and averaging over the oscillations, this translates into a difference in the stopping force of roughly 0.3~eV/\AA. Knowing that the electronic stopping force at the channel center is about 2.8 eV/$\angstrom$ ($v$ = 0.3 a.u., \cite{ponomareva2024local}), this estimate suggests a possible reduction in the ion range of the order of 0.3/2.8 $\approx$ 11\%. Figure \ref{FIG:A1}d further shows that the electron densities sampled by deeply penetrating ions differ substantially from the densities associated with idealized straight-line channeling trajectories used in the TDDFT calculations.

\begin{figure}[t!]
	\centering
	\includegraphics[width=1\columnwidth]{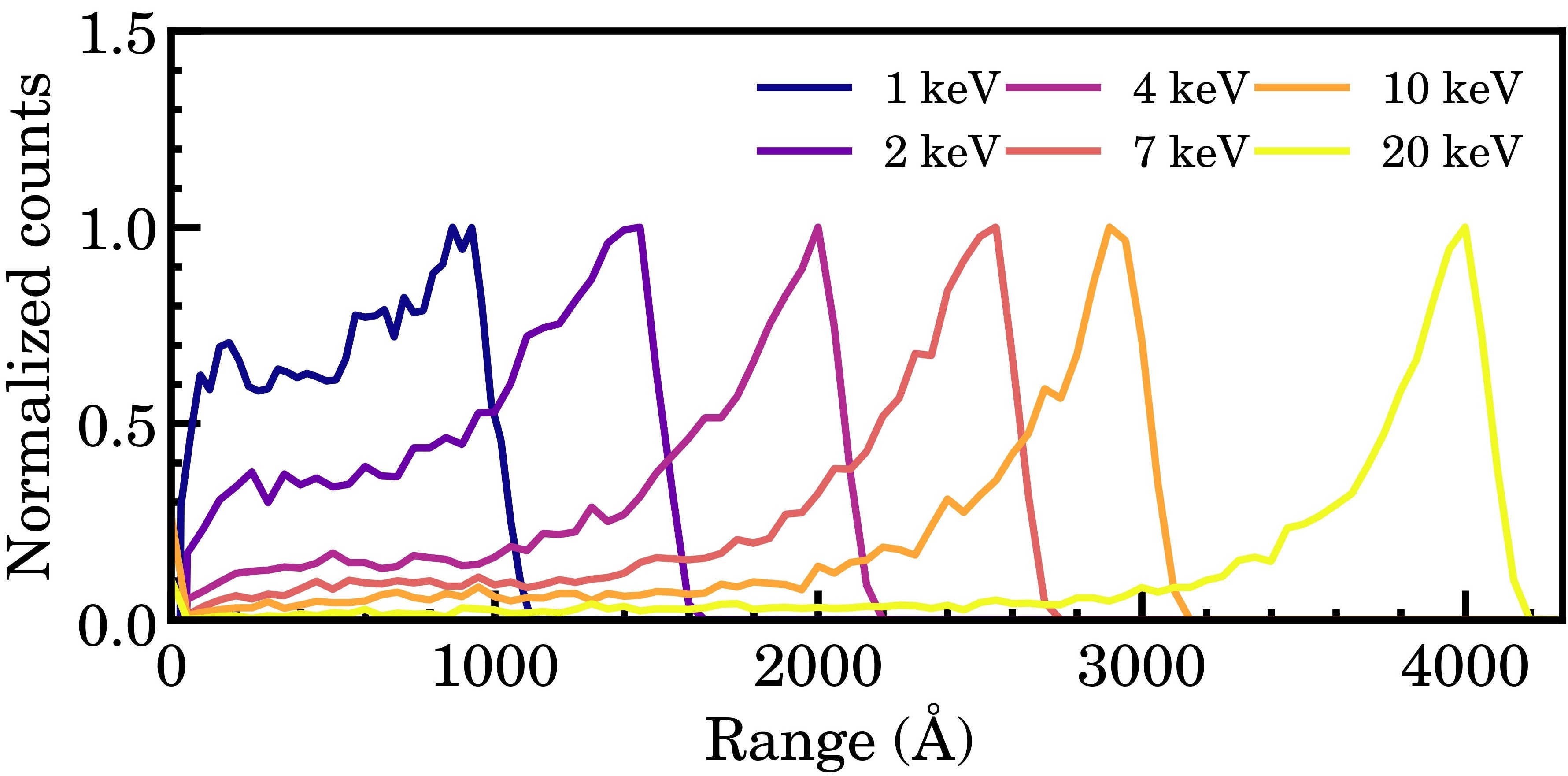}
	\caption{Calculated ion range distributions for different initial energies of H projectile moving along the $\langle100\rangle$ channel in W.}
	\label{FIG:4b}
\end{figure}

Finally, Fig. \ref{FIG:4b} presents statistical ion range profiles for the same channeling direction for H. At higher energies, the range profile exhibits a distinct peak characteristic of well-channeled ions. As the energy decreases to 1 keV, this peak broadens and gradually evolves into a double-peak structure, indicating that a significant fraction of ions undergo dechanneling and begin to traverse alternative paths within the lattice. Below 1 keV (not shown in the plots), the range distribution peak shifts toward much shorter penetration depths, signifying that most ions no longer remain channeled but instead propagate along quasi-random directions within the crystal.

This analysis indicates that the capability of an electronic stopping model to resolve local electron density effects is crucial for correctly predicting the energy-dependent channeling and dechanneling behavior and range profiles of light ions.

\section*{Appendix B. Coverage of local atomic environments by the rt-TDDFT fitting dataset}

\renewcommand{\thefigure}{B.\arabic{figure}}
\setcounter{figure}{0}

To assess how well the rt-TDDFT fitting dataset represents the local atomic environments encountered during realistic ion dynamics, we analyzed the distribution of the closest projectile–target atom distances sampled during statistical MDRANGE simulations. The analysis was performed for D backscattering from W (see Sec. 4  and Fig. \ref{FIG:8} of the main text) by averaging over 100 simulated trajectories.

\begin{figure*}[t!]
	\centering
	\includegraphics[width=0.85\textwidth]{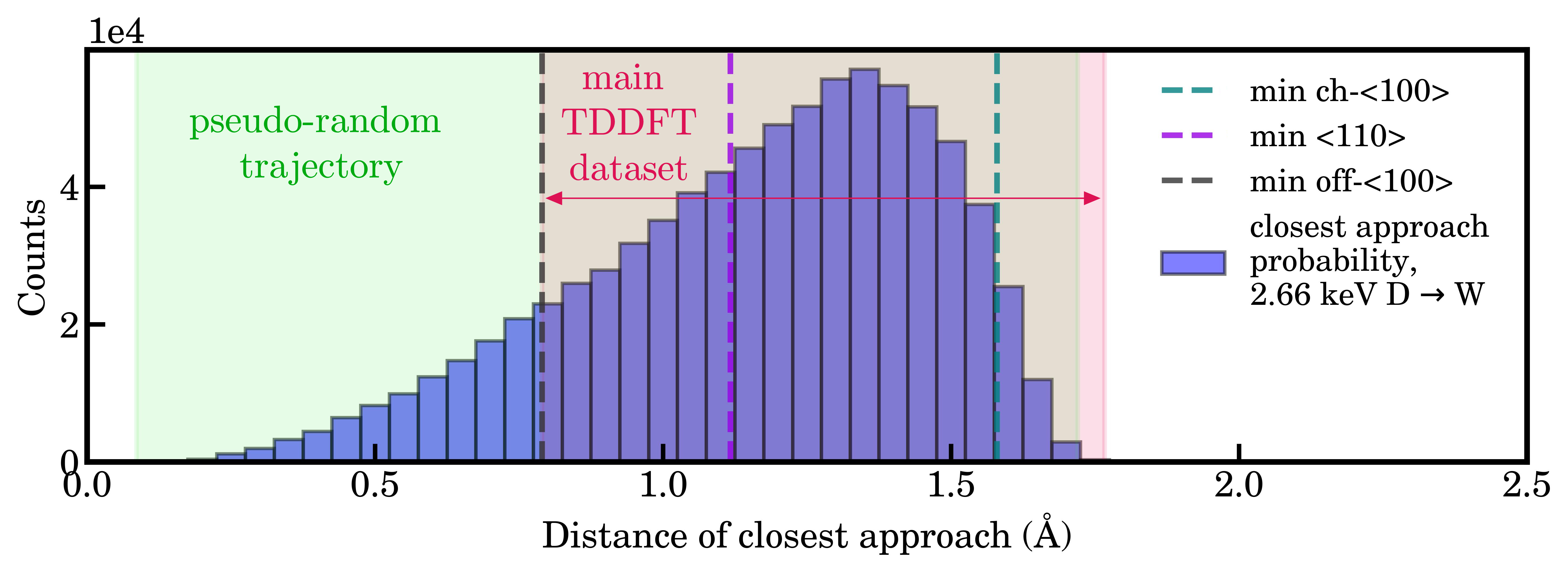}
	\caption{Distribution of the closest projectile-target atom distances sampled during MDRANGE simulations of D backscattering from W, averaged over 100 trajectories. The shaded regions indicate the range of closest approach distances covered by the rt-TDDFT fitting dataset. The dashed lines show the minimum closest approach distance along each channeling trajectory.}
	\label{FIG:B}
\end{figure*}

Figure \ref{FIG:B} compares this distribution with the range of closest approach distances covered by the rt-TDDFT trajectories used for the parameterizations. The main fitting dataset, including the channeling and off-center channeling trajectories described in Sec. 2, spans the majority of the distances sampled during the backscattering simulations. An additional pseudo-random trajectory extends the coverage towards the low-distance tail of the distribution associated with less frequent close-collision events.

As emphasized in the main text, including the pseudo-random trajectory produces only a minor change in the resulting density-dependent stopping functions and in the predicted ion ranges. This indicates that the parameterization is primarily constrained by the representative channeling trajectories, while the additional pseudo-random trajectory serves mainly to improve the description of the extreme close-collision events encountered only rarely during realistic ion propagation.


\end{document}